\documentclass[12pt,letterpaper]{article}

\usepackage{amsmath}
\usepackage{amsfonts}
\usepackage{amssymb}
\usepackage{graphicx}
\def\be{\begin{equation}}
\def\ee{\end{equation}}

\newcommand{\half}{{\textstyle{\frac12}}}

\begin{document}

\begin{flushright} {\footnotesize HUTP-04/A043}  \end{flushright}
\vspace{5mm}
\vspace{0.5cm}
\begin{center}

\def\thefootnote{\fnsymbol{footnote}}

{\Large \bf Perturbations in bouncing cosmologies:\\ \vspace{.2cm} dynamical attractor {\em vs} scale invariance} \\[1cm]
{\large Paolo Creminelli$^{\rm a}$, Alberto Nicolis$^{\rm a}$ and Matias Zaldarriaga$^{\rm a,b}$}
\\[0.5cm]

{\small 
\textit{$^{\rm a}$ Jefferson Physical Laboratory, \\
Harvard University, Cambridge, MA 02138, USA}} 

\vspace{.2cm}

{\small 
\textit{$^{\rm b}$ Center for Astrophysics, \\
Harvard University, Cambridge, MA 02138, USA
}}
\end{center}

\vspace{.8cm}

\hrule \vspace{0.3cm} 
{\small  \noindent \textbf{Abstract} \\[0.3cm]
\noindent
For bouncing cosmologies such as the ekpyrotic/cyclic scenarios we show that it is possible to make predictions for density 
perturbations which are independent of the details of the bouncing phase. This can be achieved, as in inflationary cosmology, 
thanks to the existence of a dynamical attractor, which makes local observables equal to the unperturbed solution up
to exponentially small terms. Assuming that the physics of the bounce is not extremely sensitive to these 
corrections, perturbations can be evolved even at non-linear level. The 
resulting spectrum is not scale invariant and thus incompatible with experimental data.
This can be explicitly shown in synchronous gauge where, contrary to what happens in the commonly used Newtonian gauge,
all perturbations remain small going towards the bounce and the existence of the attractor is manifest.

\vspace{0.5cm}  \hrule

\def\thefootnote{\arabic{footnote}}
\setcounter{footnote}{0}
\section{Introduction}
From cosmological observations we know that the current Universe is to a good approximation flat, homogeneous and isotropic on large 
scales. It is well known that in the standard Big Bang cosmology this requires an enormous amount of fine tuning on the initial 
state emerging from the Planck era. This problem is usually solved postulating a period of inflation which acts like a 
dynamical attractor for the cosmological evolution. From a generic initial condition sufficiently close to a flat, homogeneous 
and isotropic state the Universe evolves towards this symmetric configuration. 
Another logical possibility that has been explored is that a dynamical attractor is present in a contracting phase which precedes the
present expansion. After this phase, the Universe would reach a state of high curvature and then bounce and start expanding. 
Although string-inspired, the ekpyrotic/cyclic scenarios \cite{Khoury:2001wf,Steinhardt:2001st} are fully described in 
the contracting phase well before the bounce by a 4d effective field theory. The only light degree of freedom (besides the graviton) 
is a scalar field which moves along a steep negative potential and drives the cosmic contraction \footnote{In the pre-Big Bang 
scenario \cite{Gasperini:2002bn} the (Einstein frame) 
contraction is driven by the kinetic energy of the dilaton, which has no potential. As we will discuss this can be seen as a 
limiting case of the ekpyrotic/cyclic models and the conclusions we will draw are different.}. 
Even in the presence of perturbations, if the potential is steep enough, inhomogeneities and anisotropies become less and less 
relevant going towards the bounce and the cosmic evolution gets closer and closer to the unperturbed one. 

The crucial question is whether this class of models can give rise to an approximately scale invariant spectrum of adiabatic 
density perturbations as clearly required by experiments. The answer to this question is not straightforward because there is
no explicit model of a bouncing phase; it is just assumed that at sufficiently high energy UV corrections will stop the 
contraction and lead to an expanding phase. It is therefore important to understand if one can make predictions which 
are robust, {\em i.e.} independent of the details of the unknown bouncing phase. In inflationary cosmology
we face a closely related issue because perturbations, that are created during inflation, are observed much later and 
there are cosmological phases in between, like reheating, whose details are completely unknown. It is well
understood that in models with a single fluctuating field all predictions are independent of the details of the unknown 
cosmological phases. There are two reasons why this is true. First of all we are interested in modes which are well 
outside the horizon during the unknown regimes of the cosmological evolution. This would not be enough however. The situation 
is further simplified by the absence of isocurvature perturbations: every observer will eventually go 
through the same cosmological history. Following the dynamical attractor, after a mode goes outside the horizon, the Universe 
locally approaches the unperturbed solution up to exponentially small terms. This ``parallel Universes'' approach 
\cite{Salopek:1990jq}
allows to follow perturbations through the unknown cosmological regimes at any order in perturbation theory. On the contrary 
if we add another field into the game, separate regions of the Universe will be characterized by different values of this new
field and will thus undergo a different evolution, so that the final state depends on the details of the whole cosmological 
history. The fate of the isocurvature component can be very different: it can be washed away by thermal equilibrium or, on the 
contrary, become the leading source of adiabatic perturbations as in the curvaton and variable decay width scenarios.

The purpose of this paper is to understand whether the same kind of arguments used for single field inflation can be applied
to evolve density perturbations through the unknown bouncing phase of the ekpyrotic/cyclic scenario.

In Newtonian gauge the most generic scalar perturbation is (at linear order) a superposition of the two independent solutions for the 
Newtonian potential $\Phi$. It has been pointed out in the context of the ekpyrotic/cyclic scenario \cite{Khoury:2001zk} 
that, in the limit of a very steep potential, one of these two solutions has an approximately scale invariant spectrum in the 
contracting phase. 
Unfortunately the most naive evolution from the contracting to the expanding phase suggests that this scale invariant mode 
is orthogonal to the ``growing mode'' in the expanding phase, the one relevant for observations. However it has been
claimed that a generic mixing of the two solutions at the bounce would induce an observationally viable spectrum.
In this paper (section \ref{sec:2gauges}) we show that this generality can not be defended in any way. 
For example if we change gauge and describe the most generic perturbation in terms of the variable $\zeta$ (proportional to 
the curvature of comoving surfaces), none of the two independent modes is scale invariant. This tells us that for a contracting 
cosmology the concept of scale invariance of a particular variable has no physical meaning before one specifies how this variable 
is related to what we will eventually observe.

As we discussed, in the presence of a dynamical attractor in the contracting phase, close to the bounce 
the cosmological solution is locally ({\em i.e.}~on regions smaller than the Hubble radius) homogeneous, isotropic and flat, 
and the scalar follows its unperturbed evolution, up to exponentially small terms. This leads us to make a simple 
assumption (section \ref{sec:main}), which is implicitly made also in the case of an inflationary Universe when evolving
perturbations through unknown phases. If the physics of the bounce is not tremendously sensitive to these exponentially small 
corrections, every observer will go through the bounce in the same way, following the unperturbed evolution. This allows 
to predict, not only at linear level but at any order in the perturbation, the statistical properties of the 
fluctuations in the expanding phase. Unfortunately the conclusion is that {\em the spectrum of density fluctuations is not 
scale invariant} and thus it is ruled out by experiments. 
This can be explicitly shown in synchronous gauge where, contrary to what happens in the commonly used Newtonian gauge,
all perturbations remain small going towards the bounce and the existence of the attractor is manifest.
If our assumptions are not satisfied, which 
is a logical possibility, then all predictions, like the scale invariance of the 2-point function or the level of non-gaussianity 
of the perturbations, strongly depend on the details of the bounce and no robust prediction can be made.  

We stress that we are {\em not} giving any prescription to match perturbations across the bounce (for example we are not claiming that
$\zeta$ must be continuous \cite{Lyth:2001pf,Brandenberger:2001bs,Hwang:2001zt,Martin:2001ue}, although the final result 
for density perturbations in the expanding phase is the same).  Given a generic initial condition before the bounce 
there is no way to evolve it through the unknown phase without specifying all the details.  The point here is that
we do not have a {\em generic} initial condition but a very special one, because of the presence of
the dynamical attractor.  Locally the solution is exponentially close to the unperturbed solution and this is a powerful simplification. 
Again the situation is similar to inflationary cosmology; nobody knows how to evolve a generic initial state through reheating because we do
not know how reheating looks like, but there is no need for this as inflation leads to a very specific state, locally (but not on large
scales) indistinguishable from the unperturbed solution.

\section{Intuitive argument and its limitations}\label{intuitive}
In this section we want to approach the problem by a (critical) review of the basic heuristic argument for
a scale-invariant spectrum of density perturbations in a bouncing cosmology 
\cite{Khoury:2001zk,Steinhardt:2002kw,Steinhardt:2004gk}. This will show that there are qualitative differences with respect to an 
inflationary scenario that must be taken into account. For the time being we neglect how perturbations will evolve through 
the bounce. The argument is an estimate of the fluctuations of the scalar field $\phi$, which is driving 
the contraction of the Universe towards the bounce. As the field is pulled by a steep negative 
potential, it seems natural to assume that gravity is negligible in the dynamics of the fluctuations of $\phi$. 
Extending the discussion of refs.~\cite{Lyth:2001pf,Brandenberger:2001bs}, we will argue that this is not the case.   

To simplify the algebra it is useful to approximate the potential over some range of $\phi$ with the exponential form
\cite{Khoury:2001zk}
\be
V (\phi) = -V_0 \: e^{-\sqrt{\frac2p}\frac\phi{M_P}}\;,\qquad M_P \equiv (8 \pi G)^{-1/2}\; .
\ee
In the ekpyrotic/cyclic scenario the potential is very steep, {\em i.e.}~$p \ll 1$. The Friedmann equations and the 
equation of motion for an homogeneous $\phi$ configuration,
\be
\label{eq:phiback}
\ddot \phi + 3 H \dot \phi + V'(\phi) = 0 \; ,
\ee
are exactly solved by the background
\be
a(t) = (t/t_0)^p \; , \quad H(t) = \frac p t \;, \quad \phi_0(t) = M_P \sqrt{2p} \: \log \frac{t}{t_0} \; ,
\ee
where $t_0=-\sqrt{p(1-3p)}\cdot M_P/\sqrt{V_0}$.  The scalar $\phi$ is moving from $+\infty$ to $-\infty$ and $t$ is negative 
and running towards $0$ which corresponds to the unknown bouncing phase. This solution has a constant, large 
pressure-to-energy density ratio, $w = (2/3p) -1 \gg 1$. 

It can be shown that for $p < 1/3$ this solution is a {\em dynamical attractor}, in the sense that
the system rapidly approaches it starting from a generic homogeneous and isotropic initial condition \cite{Heard:2002dr}. 
Roughly 
speaking, in the presence of a very steep potential the kinetic energy of the scalar becomes bigger and bigger in the evolution, 
so that the asymptotic  state does not depend on the initial kinetic energy. In the following we will explicitly see that the 
same conclusion remains true in the presence of (sufficiently small) departures from isotropy and homogeneity. As discussed,
this property will turn out to be crucial for our argument.

Since in the limit $p \ll 1$ the scale factor changes very slowly with time, one could be tempted to say that the 
gravitational backreaction is small and study perturbations in the scalar $\phi$ on the fixed gravity background. The equation
of motion on a fixed FRW metric would be 
\be
\label{eq:dec}
\delta \ddot \phi_{\vec k} + 3 H \delta \dot \phi_{\vec k} + \left(\frac{k^2}{a^2} + V''(\phi)\right) \delta\phi_{\vec k} = 0 \; ,
\ee
where $\vec k$ is the comoving wavevector. For the background above we get
\be\label{eq:dec2}
\delta \ddot \phi_{\vec k} + \frac{3 p}{t}  \delta \dot \phi_{\vec k} + \left(\frac{k^2}{a^2} 
- \frac{2(1-3p)}{t^2}\right) \delta\phi_{\vec k} = 0 \;. 
\ee
It is evident that the contraction of the Universe gives a subleading contribution for $p \ll 1$. 
Moreover for small $t$ the gradient term can be neglected so that the mode freezes. 
The equation reduces to $\delta \ddot \phi_{\vec k} - \frac2{t^2} \delta\phi_{\vec k} = 0$, which gives the two behaviors 
$\delta\phi_{\vec k} \sim t^{-1}$ and $t^2$. 
Therefore the leading solution for small $t$, properly normalized to match the usual Minkowski limit for large negative $t$, is 
\be
\delta\phi_{\vec k} \simeq \frac1{\sqrt{k}} \cdot \frac1{k t} \;.
\ee
This implies a scale-invariant spectrum in position space as 
$\langle\delta\phi(x)^2\rangle \sim \int \!d \log k \; k^3 | \delta\phi_{\vec k}| ^2 $. 

It is however easy to realize that the argument above is not physically motivated, even at the heuristic level. 
In fact there is no limit in which the fluctuations of $\phi$ can be decoupled from the metric perturbations since $\phi$ 
is what drives the evolution of the Universe and its potential is steep. 
Taking the naive decoupling limit $M_P \rightarrow \infty$ one obviously ends up with a trivial background. The only correct decoupling 
limit would be to have a flat potential for $\phi$; in this case eq.~(\ref{eq:dec}) is correct because fluctuations of $\phi$ do not 
gravitate except for gradient terms. This is the reason why in slow-roll inflation the intuitive argument above gives a correct 
estimate of the density perturbations up to slow-roll corrections.   

In our case there is no sense in which the dynamics of perturbations is dominated by the potential term: gravity is a crucial
ingredient and must be taken into account. Let us clarify this statement in a gauge which is quite close to the intuition of 
keeping $\delta\phi$ as the dynamical variable characterizing scalar perturbations,
\be \label{gauge_deltaphi}
\delta\phi(\vec x,t) =\phi(\vec x,t) -\phi_0(t) \; , \qquad g_{ij} = a^2(t) \delta_{ij} \;.
\ee
Unless otherwise specified we concentrate to scalar modes, setting tensor modes to zero.
In this gauge gravitational effects are suppressed in inflation by slow-roll parameters, so that
it explicitly realizes the decoupling limit discussed above.

The other components of the metric, $g_{00}$ and $g_{0i}$, are Lagrange multipliers and as such are determined as a function of 
$\delta\phi$ through the constraint Einstein equations. Substituting them into the action for gravity and the scalar field and
specializing to our background solution we get a Lagrangian for $\delta\phi$ of the form \cite{Maldacena:2002vr}
(see Appendix \ref{app_action} for the derivation of this result)
\be \label{action_deltaphi}
S = -\half \int \!d^3x dt \sqrt{-g} \;(\partial_\mu\,\delta\phi)^2 \;.
\ee
This is the Lagrangian for a scalar in the given background but {\em without any potential term}, so that from this point of view 
it is difficult to argue that the potential is driving the dynamics of the perturbations. Of course the equation
of motion deriving from this Lagrangian is exactly eq.~(\ref{eq:phiback}) without the mass term. The effect of metric perturbations is 
so important that exactly cancels out the potential.
Note that the very simple Lagrangian above is exact only in a scaling solution $a(t) \propto t^p$, like the collapsing 
background described above ($p \ll 1$) or inflation with an exponential potential ($p \gg 1$). In both cases the action will receive
small corrections proportional to the variation of the equation of state in one Hubble time, {\em i.e.}~subleading respectively
in the fast-roll or slow-roll expansion.

From the action above, the properly normalized solutions for small $t$ are now $\delta\phi_{\vec k} \sim 1/\sqrt k$ and 
$\delta\phi_{\vec k} \sim \sqrt k \: t$. None of these implies a scale-invariant spectrum. 

To get further insight into the problem, in the next section we will perform a detailed analysis of scalar perturbations in 
two different gauges.

\section{Standard gauges and the apparent rapid growth of perturbations}\label{sec:2gauges}
The study of scalar perturbations can be performed in different gauges. In each one, after satisfying the constraint
equations, the perturbation is parametrized by a single scalar function, which satisfies, at linear order,
a second order linear differential equation. The most generic scalar fluctuation will therefore be described
by a linear combination of the two independent solutions of this equation. 
Usually one of the solutions dominates over the other at late times (``growing''and ``decaying'' mode). 
However in some cases the same physical perturbation is described in one gauge as growing mode 
and in another as decaying (we will see an example of this in the following).
Neglecting the decaying mode is therefore quite misleading, 
especially in a contracting background like the one at hand, and thus will be avoided. 

To make contact with the literature, in this section we are going to perform the calculation in two commonly used gauges. 
Doing so we will stress some subtle point which has been usually overlooked. However it will turn out that in both gauges perturbation 
theory breaks down for $t \to 0$ in the background under study. In the next section we will therefore move to synchronous gauge in 
which the perturbativity of the calculation remains manifest.

The analysis of perturbations in inflation is greatly simplified by using the variable $\zeta$ introduced in 
ref.~\cite{Bardeen:1983qw}, because it is conserved outside the horizon under simple assumptions on the stress energy
tensor. We therefore start our analysis in the $\zeta$-gauge defined by
\be
\phi(\vec x,t) = \phi_0(t) \; , \qquad g_{ij} = (1+2 \zeta(\vec x,t)) \: a^2(t) \delta_{ij} \;.
\ee
In this gauge the scalar field is unperturbed so that the relation with 
the gauge introduced in the previous section is simply a time reparameterization
\be
t \to t - \frac{\delta\phi}{\dot\phi_0} \qquad \Rightarrow \qquad \zeta = - \frac{H}{\dot\phi_0}\delta\phi \;.
\ee
In our scaling background solution $\delta\phi$ and $\zeta$ are simply related by a constant factor
\be
\zeta = - \sqrt{\frac p 2}\frac1{M_P} \delta\phi
\ee
so that the Lagrangian for $\zeta$ is again very simple
\be \label{action_zeta}
S = -\frac{M_P^2}{p}\int \!d^3x dt \sqrt{-g} \;(\partial_\mu\,\zeta)^2 \;.
\ee
The other components of the metric are fixed as functions of $\zeta$ through the constraint equations. 
At linear order they are (see Appendix \ref{app_action} for details)
\begin{eqnarray}
g_{00} & = & -1 - 2 \frac{\dot\zeta}{H} = -1 - \frac{2 t}{p} \dot\zeta \\ \label{eq:offd}
g_{0i} & = & - \partial_i \left(\frac{\zeta}{H} -  \frac{\dot\phi_0^2}{2H^2M_P^2}\,\frac{a^2}{\partial^2}\dot\zeta \right) 
= - \partial_i \left(\frac{t}{p}\zeta - \frac{1}{p}\frac{(t/t_0)^{2p}}{\partial^2}  \dot\zeta \right) \;,
\end{eqnarray}
where, as usual, $1/\partial^2$ is defined in Fourier transform.
Note that $\zeta$ is related to the Ricci scalar of the spatial metric induced on comoving surfaces ({\em i.e.}~spatial
hypersurfaces of constant $\phi$) simply by
\be \label{eq:3R}
^{(3)}R = 4 \frac{k^2}{a^2} \zeta_{\vec k} \; ,
\ee
as in this gauge comoving surfaces are surfaces of constant $t$.

The equation of motion for $\zeta$ deriving from its action above is
\be\label{eq:zeta}
\ddot\zeta_{\vec k} + \frac{3p}{t} \dot\zeta_{\vec k} + \frac{k^2}{(t/t_0)^{2p}} \zeta_{\vec k} = 0 \;,
\ee
which can be integrated in terms of $J$ and $Y$ Bessel functions with argument $k\tau$, where $\tau$ is the conformal time defined
as usual by $dt = a(t) d\tau$. 
The full solution is
\be
\zeta_{\vec k}(\tau) = \frac{1}{a} \sqrt{-k \tau} \left( A_{\vec k} \, J_{\frac{1-3p}{2-2p}}(-k\tau) + 
B_{\vec k} \, Y_{\frac{1-3p}{2-2p}}(-k\tau)\right) \; ,
\ee
where $A_{\vec k}$ and $B_{\vec k}$ are integration constants, to be determined below.
It has been claimed that the variable $\zeta$ is not suitable for discussing density perturbations in a contracting phase,
because it shows no ``classical instability'' and therefore is not ``amplified'' \cite{Lyth:2001pf,Khoury:2001zk}. The meaning 
of this is unclear since the two modes for $\zeta$ above parametrize the most generic scalar perturbation of the system.

Following the standard procedure \cite{Birrell:1982ix}, we then quantize the system by writing the quantum field operator as
$\hat \zeta_{\vec k}(\tau) = \zeta_{\vec k}(\tau) \hat a_{\vec k} + \zeta^*_{\vec k}(\tau) \hat a^\dagger_{-\vec k}$,
where $\zeta_{\vec k}(\tau)$ is the above classical solution, and the creation-annihilation operators obey
the standard commutation rule $[\hat a_{\vec k}, \hat a^\dagger_{\vec k'}] = (2\pi)^3 \delta^3(\vec k-\vec k')$.
Since for large negative $\tau$ the Universe reduces to Minkowski spacetime, we impose that the classical solution
$\zeta_{\vec k}(\tau)$ asymptotes to the standard Minkowski wavefunction $\propto\frac{1}{\sqrt{2 k}} e^{-i \, k \tau}$
of a massless scalar field. Also, we assume that the initial state is the usual Minkowski vacuum.
The proper normalization can be read from the action eq.~(\ref{action_zeta}) written in conformal time,
\be
\zeta_{\vec k}(\tau) \to \frac{\sqrt{p}}{\sqrt 2 M_P}\frac{1}{a} \cdot \frac{1}{\sqrt{2 k}} e^{-i \, k \tau}  \qquad 
\mbox{for} \quad \tau \to -\infty \; .
\ee
This fixes the linear combination of the two independent solutions to be
\be\label{eq:Bessel}
\zeta_{\vec k}(\tau) = \frac{1}{a} \frac{1}{\sqrt{2 k}} \frac{\sqrt{\pi}}{2 M_P}\cdot \sqrt{p}
\sqrt{-k \tau} \left( J_{\frac{1-3p}{2-2p}}(-k\tau) + i\,  Y_{\frac{1-3p}{2-2p}}(-k\tau)\right) \; .
\ee
At small $t$ we recover the two behaviors that solve eq.~(\ref{eq:zeta}) neglecting gradient terms 
\be \label{eq:zetabh}
\zeta_{\vec k} \sim \frac1{M_P} (i k^{-\frac12 +p} t_0^p + k^{\frac12 -p} t^{1-3p} t_0^{2p}) \; ,
\ee
where we expanded the exponents up to linear order in $p$.
The first term comes from the $Y$ function while the second comes, for generic $p$, both from $Y$ and $J$. Even though
$\zeta$ remains finite for $t \to 0$ the off-diagonal term $g_{0i}$ in eq.~(\ref{eq:offd}) diverges like $t^{-p}$, so that linear
theory cannot be justified in this gauge for small $t$.

The late time spectrum of $\zeta$ is 
\be
\langle \hat \zeta_{\vec k}(t) \hat \zeta_{\vec k'}(t) \rangle = (2\pi)^3 \delta^3(\vec k+\vec k') \:
|\zeta_{\vec k}(t)|^2 \; , 
\ee
with
\be\label{eq:spectrum}
|\zeta_{\vec k}(t)|^2 \sim 
\frac1{M_P^2} (k^{-1 + 2p} t_0^{2p} + k^{1 -2p} t^{2-6p} t_0^{4p}) \; ,
\ee
which does not contain any scale-invariant component. This remains true even if one includes all further terms in the 
$k\tau$ expansion of the complete solution eq.~(\ref{eq:Bessel}) as they only add positive powers of $k$. This result can be
written in a more transparent form as a function of the physical wavevector $k_{\rm ph}=k/a$,
\be \label{zeta_physical}
\langle \zeta(x)^2 \rangle \sim \int\! d \log k_{\rm ph} \;\; p \frac{H^2}{M_P^2} \left[\left(k_{\rm ph}/H\right)^{2+2p} 
+ \left(k_{\rm ph}/ H\right)^{4-2p} \right] \;.
\ee

We stress again that the complete solution for $\zeta$ describes the most generic scalar fluctuation
of the system. But, as discussed in the Introduction, a scale invariant spectrum is obtained for the Newtonian potential 
$\Phi$  so that the situation is at first puzzling.

Before analyzing the relation between the results obtained for $\zeta$ and $\Phi$, it is important to appreciate
that the second term in eq.~(\ref{eq:zetabh}) cannot be simply disregarded as ``decaying''.
Although it drops off in time like $t^{1-3p}$, thus giving a negligible contribution to the intrinsic
curvature of comoving surfaces (see eq.~(\ref{eq:3R})), it dominates the perturbation in the {\em extrinsic} curvature
of comoving surfaces.
In fact at linear order the extrinsic curvature is
\be\label{eq:zetaex}
K^i{}_j = \half g^{ik}( \dot g_{kj} - \partial_k g_{0j} - \partial_j g_{k0}) = 
(H +\dot\zeta) \delta^i_j -\frac1p \frac{\partial_i \partial_j}{\partial^2}  \dot\zeta  - 
\frac tp \frac{\partial_i \partial_j}{a^2}  \zeta\;.
\ee
The leading contribution comes from the term $\zeta_{\vec k} \sim t^{1-3p}$ in eq.~(\ref{eq:zetabh}), which gives a perturbation in 
$K^i{}_j$ diverging as $t^{-3p}$ towards the bounce. Note that this is however a small perturbation with respect to the unperturbed 
extrinsic curvature which diverges faster, as $H \sim t^{-1}$.
We see that the two terms in eq.~(\ref{eq:zetabh}) are both physically relevant as they dominate respectively the intrinsic 
and extrinsic curvature. Further terms in the small $t$ expansion of eq.~(\ref{eq:Bessel}) are instead subleading in their
contribution to physical curvatures. 

Let us now see how these perturbations are reinterpreted in the commonly employed Newtonian gauge. This gauge is defined by
\be
ds^2 = -(1 + 2 \Phi) dt^2 + (1-2\Psi) a^2(t) d\vec{x}^2 \;.
\ee
The most generic scalar perturbation can be described by the Newtonian potential $\Phi$. All other variables are related to $\Phi$ 
through the constraint Einstein equations. In particular $\Psi = \Phi$ at linear order in absence of anisotropic 
stress, as in our case. The equation of motion for $\Phi$ in our background is (see for example ref.~\cite{Khoury:2001zk})
\be
\ddot\Phi_{\vec k} + \frac{2+p}{t} \dot\Phi_{\vec k} + \frac{k^2}{(t/t_0)^{2p}} \Phi_{\vec k} = 0 \;.
\ee
Following the usual procedure as done for $\zeta$ we get the properly normalized solution
\be
\Phi_{\vec k} (\tau) = \frac{1}{a} \frac{1}{\sqrt{2  k}}  \frac{\sqrt{\pi}}{2 M_P}\cdot \frac{p}{1-p}
\frac{1}{\sqrt{-k \tau}} \left( J_{\frac{1+p}{2-2p}}(-k\tau) + i\,  Y_{\frac{1+p}{2-2p}}(-k\tau)\right) \;,
\ee
with late time behavior 
\be
\label{eq:philate}
\Phi_{\vec k} \sim \frac1{M_P} (k^{-3/2-p} t^{-1-p} + k^{-1/2+p} t_0^p) \;.
\ee
We see that the first term, which diverges approaching the bounce (thus casting doubt on the linear 
approximation), has an approximately scale invariant dependence on $k$ 
as first pointed out in ref.~\cite{Khoury:2001zk} (\footnote{Note that even though this seems to validate the naive
argument discussed in the previous section, the time dependence of the second solution, $\Phi =$ const, is not a solution
of eq.~(\ref{eq:dec2}). The same happens for $\delta\phi$ in Newtonian gauge which has the same time dependence as $\Phi$ 
(see eq.~(\ref{eq:dphi})).}).  
But how is it possible that here we have an almost scale invariant spectrum, while there is no sign of scale-invariance in 
$\zeta$? After all, we said that the most generic perturbation can be described using either $\zeta$ or $\Phi$. 
It is straightforward to verify that the scale invariant term in the expression for $\Phi$ dominates the extrinsic 
curvature of comoving surfaces. In fact, from the relation between $\delta\phi$ and $\Phi$ in this gauge 
\cite{Khoury:2001zk} 
\be\label{eq:dphi}
\delta\phi = \frac{2 M_P^2}{\dot\phi_0}(\dot\Phi + H \Phi) \;,
\ee
we obtain the contribution of the first term in eq.~(\ref{eq:philate}) to the extrinsic curvature of the $\delta\phi =0$ 
surfaces 
\be\label{eq:Phiex}
K^i{}_j \sim H^{-1} \frac{k_i k_j}{a^2} \Phi_{\vec k} \propto t^{-3p} k^{1/2} \; ,
\ee
which is the same $t$ and $k$ dependence we obtained in the previous gauge, eq.~(\ref{eq:zetaex}) 
(\footnote{To be precise only the anisotropic piece of the extrinsic curvature in eq.~(\ref{eq:zetaex}) is recovered in this
way. To get the isotropic component proportional to $\dot\zeta$ we have to keep further terms ($\sim t^{1-3p}$) in the small 
$t$ expansion of $\Phi$ in addition to those of eq.~(\ref{eq:philate}). This again tells us that it is not obvious to get the physical 
importance of a term by simply looking at its contribution to $\Phi$; a term diverging as $t^{-1-p}$ and one decaying as 
$t^{1-3p}$ give comparable extrinsic curvatures!}). 
Notice that 
the different dependence on $k$ of the two variables $\zeta$ and $\Phi$ corresponds to the fact that their relation with physical 
quantities like curvatures can involve a different number of derivatives. 
{\em This tells us that the scale invariance of a scalar fluctuation has no well defined meaning before specifying which 
variable is relevant for observation}.
The scale-invariant term in eq.~(\ref{eq:philate}) gives a vanishing contribution to the intrinsic curvature eq.~(\ref{eq:3R}).
This is clear from the explicit relation between $\zeta$ and $\Phi$
\be
\zeta = \frac{2}{3 (1+w)} \cdot \frac1a\frac{d}{d t}\left(\frac{\Phi}{H/a}\right) \;,
\ee
that is why there is no sign of scale-invariance in $\zeta$. The fact that $\zeta$ vanishes for the first term in eq.~(\ref{eq:philate})
could suggest that we are losing it in going to the variable $\zeta$, but as we saw its geometric effect in perturbing the
extrinsic curvature is completely captured by $\zeta$. This is also a nice example of how misleading the description
of a solution as decaying or growing can be; the behavior $\Phi \sim t^{-1-p}$ would be considered as growing in Newtonian gauge, but
we saw that the same physical fluctuation is ``decaying'' in $\zeta$ gauge, $\zeta \sim t^{1-3p}$.

\section{Synchronous gauge: making the dynamical attractor manifest}\label{sec:main}
We are now going to discuss the relation between the scalar perturbations produced in the contracting phase and what is 
observed today. 
The behavior at late times of scalar fluctuations in an expanding phase can be obtained from our eqs.~(\ref{eq:zetabh}) 
and (\ref{eq:philate}) (with generic coefficients), remembering that now $t$ is positive and going towards $+\infty$. Note that 
now the time independent piece in both $\Phi$ and $\zeta$ dominates each variable, with the time dependent piece going to
zero (now $p = 2/3$ in matter dominance and $p = 1/2$ in radiation dominance). 

In the present expanding phase all the ambiguity described above about the different intuition in the two gauges and 
the distinction between growing and decaying modes is absent. This is because once the fluctuation reenters the horizon 
all physical quantities depend (apart from exponentially small corrections) only on the 
$\zeta \sim \Phi \sim$ const mode; the concept of growing and decaying modes has thus a clear physical meaning. To have a viable
model the constant mode must have an approximately scale invariant spectrum. Even though the time independent term of $\Phi$ in 
eq.~(\ref{eq:philate}) does not have a scale invariant spectrum in the contracting phase, it has been argued that the 
bounce can induce a generic ``mixing'' between the two solutions, so that the scale invariant spectrum of the 
$\Phi \sim t^{-1-p}$ piece before the bounce is inherited also by the $\Phi =$ const solution in the expanding phase. 
It is however clear from the discussion above that it is difficult to defend this point of view, as in the $\zeta$-gauge 
there is no term with a scale invariant spectrum. It is therefore crucial to identify the relevant variable in which 
this mixing, if any, happens.

The first issue one should address in studying the evolution of the perturbations towards the bounce is the validity
of perturbation theory. In Newtonian gauge both $\Phi$ and $\delta\phi$ blow up as $t^{-1-p}$; in the $\zeta$-gauge, 
even though $\zeta$ remains finite all the way to the bounce, the $g_{0i}$ components of the metric eq.~(\ref{eq:offd}) diverge 
as $t^{-p}$. 
These facts indicate that a fully non-linear description might be required to approach the bounce in these gauges. Note however 
that as $t \to 0$ the physical curvatures induced by the perturbations are smaller and smaller with respect to $H$, 
which sets the scale of the unperturbed curvatures. Similarly to the inflationary scenario, the Universe becomes closer and 
closer to the unperturbed solution, as implied by the existence of an attractor. Thus, it is obviously useful to look for a 
gauge in which the smallness of perturbations remains manifest approaching the bounce.

This is achieved in synchronous gauge defined by $g_{00} = -1$, $g_{0i} = 0$.
Performing a gauge transformation from the $\zeta$-gauge and specializing to our scaling background we get for the metric
and the scalar field\footnote{The synchronous gauge does not fix completely the reparameterization invariance and one 
should get rid of the remaining gauge modes when writing the equations of motion in this gauge. However the solutions for $\zeta$ are 
here derived in a different gauge.}
\begin{eqnarray}
\label{eq:syncg}
ds^2 & = & -dt^2 + a^2(t) \left[\Big(1+ 2\zeta - 2 H \!\int_0^t\!dt'\,\frac{\dot\zeta}{H}\Big) \delta_{ij} - 
\frac2p \frac{\partial_i \partial_j}{\partial^2} \zeta + {\cal{O}}\big(\zeta \cdot k^2 t^{2-2p}\big)\right] dx^i dx^j \\ 
\label{eq:syncphi} \delta\phi & = & -\frac{\sqrt{2p} M_P}{t} \int_0^t\! dt' \,\frac{\dot\zeta}{H}\;.
\end{eqnarray}
These expressions are explicitly derived in Appendix \ref{app_gauge}. 
In the anisotropic piece of the metric only the time dependent part of $\zeta$ is relevant as a constant term can be set to 
zero by a spatial gauge transformation. 
The terms in brackets are either constant or decaying in time, making manifest that perturbations stay small and we are 
approaching the unperturbed solution; perturbations are becoming locally unobservable. After a mode goes outside the horizon
a local observer will still feel a residual curvature, with a consequent deviation from the unperturbed FRW solution. A curvature
term blueshifts as $a^{-2} \sim t^{-2p}$, so that it becomes rapidly irrelevant with respect to $H^2 \sim t^{-2}$; this 
effect is described by terms going as $t^{2-2p}$ in the solution above. In the metric above there are also terms
decaying as $t^{1-3p}$, {\em i.e.} slower than curvature terms. They describe the effect of anisotropies, which
blueshift as $a^{-6}$. Their time dependence is compatible with the fact that they become 
irrelevant as long as the unperturbed solution has $w > 1$ (so that its energy density blueshifts faster than $a^{-6}$) 
as we are assuming. The physical significance of the 
terms that we have sketched will be much more transparent when we will describe the full non-linear solution close to the 
bounce in the next section. Finally also $\delta\phi$ goes to zero in this gauge, as $t^{1-3p}$; again 
this agrees with the existence of an attractor for $p < 1/3$. This shows that the divergence of the Newtonian potential is 
just a gauge artifact; the deviations from the unperturbed solution are getting small.

We now have a good description of the perturbed metric before the bounce, in which linear theory does not break down. 
We can finally address the issue of how perturbations evolve through the bounce to the expanding phase. We are not
going to specify anything about the transition between the contracting and expanding regimes; our purpose is to 
understand at what level the final predictions are independent of the details of the bounce. The crucial 
simplification comes from the fact that we are interested only in modes with huge wavelength compared to 
the Hubble radius before we enter in the unknown bouncing phase. From the expression above the contribution to any local
observable ${\cal{O}}$ of a given mode with physical wavevector $k_{\rm ph}$ is suppressed with respect to the time-independent 
part of $\zeta$ (which we call $\zeta^{\rm const}$) by  
\be
\frac{\Delta {\cal{O}}}{\cal{O}} \sim \zeta^{\rm const} \left(\frac{k_{\rm ph}}{H}\right)^{1-2p} \;.
\ee
If we assume that the steep potential phase lasts at least 60 $e$-folds, we obtain that the modes of interest have 
an incredibly small effect for an observer entering the bouncing phase. The ratio $(k_{\rm ph}/H)^{1-2p}$ is of order $e^{-60}$
and it describes the fact that the local effect of a mode is very small in the presence of the dynamical attractor. Moreover the 
initial departures from the unperturbed solution when the mode leaves the horizon are of order 
$\sqrt{p} \cdot H_{\rm crossing}/M_P$ (see eq.~(\ref{zeta_physical})). This gives the size of $\zeta^{\rm const}$ in the equation above. 
For the modes of interest this gives a further exponential suppression $e^{-60}$. 

In the concrete proposed models the scaling solution we are describing ceases to 
be a good approximation even before we enter the bouncing phase, because the potential is modified at large negative $\phi$. 
The scaling solution is however valid when all cosmologically relevant modes exit the horizon; this ensures that their physical
local effect gets exponentially small.  

These considerations lead us to a simple assumption, which makes the predictions of this class of models independent 
of the details of the bounce. We assume that the unknown phase is not sensitive to this very tiny effects; {\em every
comoving observer goes through exactly the same history (whatever it is) apart from exponentially small deviations}. 
With this assumption we can neglect all terms going to zero in the metric above
\be\label{eq:onlyzeta}
ds^2 = -dt^2 + a^2(t) (1+ 2\zeta^{\rm const}(\vec x)) d\vec x^2 \;,
\ee
where $\zeta^{\rm const}(\vec x)$ is the time-independent term of $\zeta(\vec x, t)$. It is important to realize that 
$\zeta^{\rm const}(\vec x)$ has an exponentially small local effect as, neglecting gradients, it can be locally reabsorbed with a 
rescaling of the spatial coordinates. Now the crucial point is that the form of the metric eq.~(\ref{eq:onlyzeta}) will 
hold also in the expanding phase after the bounce, with the same time-independent rescaling of the spatial coordinates 
$\zeta^{\rm const}(\vec x)$. Assuming that we have a continuous history through the bounce, any observer is going through the 
same unperturbed history ($H(t)$, $\phi(t)$) once we have neglected the exponentially small corrections. But at each point 
we have in the metric eq.~(\ref{eq:onlyzeta}) a different normalization of the scale factor $(1+ \zeta^{\rm const}(\vec x)) a(t)$. 
This implies that $\zeta^{\rm const}(\vec x)$ cannot change as it is locally an integration constant of the unperturbed solution.
The reader could be worried that our conclusions depend on the validity of a 4d metric description of the bounce with
a continuous solution for the scale factor $a(t)$. Actually it is easy to realize that the result holds 
independently of the unknown physics of the bounce, even if it cannot be described in a 4-dimensional field theory language.
Given an unperturbed solution, every comoving observer will follow it (apart from exponentially small deviations), so that
the measured red-shift factor between a given comoving time $t$ before the bounce and a given time after must be the same for
all observers; $\zeta^{\rm const}(\vec x)$ therefore cannot be modified in the unknown phase.

Our assumption therefore implies that the time independent term of $\zeta$ is the same before and after the bounce. The 
spectrum of density perturbations at horizon re-entering $(\delta\rho/\rho)_*$ in the present expanding phase is thus 
given by the first term in 
eq.~(\ref{eq:spectrum}),
\be
\langle\left(\delta\rho/\rho\right)_*^2\rangle \sim \langle\zeta\zeta\rangle \propto k^{-1+2p} \;.
\ee 
The spectrum is not scale invariant and therefore incompatible with observations.

We want to stress that our assumption is not trivial. The Hubble constant must change sign during the bounce; 
one can therefore envisage a model in which the bounce is sufficiently ``slow'' so that $H$ remains small for a while and all 
interesting modes come back into the horizon and start oscillating before freezing again \cite{Gordon:2002jw}. 
From all we said it should be clear that {\em whether or not we obtain a scale invariant spectrum does now depend on the 
unknown physics at the bounce and it is not a generic prediction}. For example we saw that $\zeta$ contains no scale 
invariant term, so that it is not possible, without specifying the model, to state that a generic mixing of modes 
will give a viable prediction. 

Another possibility to get around our conclusion without allowing all the modes to come back into the horizon is to assume
that the physics at the bounce is sensitive to the exponentially small corrections we have neglected. The local
effect of the perturbations, which has been exponentially suppressed during the contraction as a consequence of the attractor
behavior, could be amplified and ``resurrected''. This is physically hard to believe and it would require a certain amount 
of fine tuning to pump up a tremendously small effect and stop at the desired $10^{-5}$ level before entering the non-linear 
regime. Anyway also this possibility does not generically give a scale invariant spectrum.

To avoid confusion it is important to underline that our approach does not tell us how to evolve a {\em generic} scalar 
perturbation to the expanding phase. A generic fluctuation will be characterized by two constants associated
with the two independent modes. Different comoving observers will follow different histories, for example at the same
value of $H$ they will see a different value of the scalar field. How the bounce occurs will depend on these differences
and to know the final state we must obviously specify the details of the bounce. Fortunately the existence of an
attractor solution drives every observer to see the same history, which is locally indistinguishable from the unperturbed one:
at the same value of $H$ we have the same value of $\phi$ and $\dot\phi$ apart from exponentially small terms. With this
crucial simplification the evolution of perturbations through the bounce is fixed by the unperturbed solution. 

An approximately scale invariant 2-point function is not the only constrain a model of the early Universe must satisfy. 
Present data constrain the non-gaussianity of the perturbations to be below $10^{-3}$ \cite{Komatsu:2003fd} 
and this also needs to be explained.
In the next section we extend our procedure to the non-linear case, so that perturbations can be followed to our 
present expanding phase also at non-linear level. If our assumption of neglecting the exponentially small terms is 
disregarded, the predictions for the higher moments of the perturbations will again depend on the unknown physics at the bounce.

\section{Non-linearly through the bounce}
Given our assumptions, we have shown in the last section that a bouncing model is not compatible with data. It seems therefore
useless to proceed with our analysis at the non-linear level. We do it for several reasons. First of all, it clarifies the
physical meaning of our argument, making it clear that it is not based on any linear approximation. Second it shows
the analogy with the similar problem of following perturbations at the non-linear level in inflation, for example to calculate
the higher moments of the spectrum. Finally it is interesting to understand how predictions can be made even if the 
cosmological evolution goes through a high curvature and strong coupling regime at the bounce, which at first sight seems to jeopardize 
any perturbative approach.

We have already stressed many times that after a mode leaves the horizon its effect becomes rapidly negligible for a local 
observer, analogously to what happens in inflation. A perturbation induces an intrinsic curvature of comoving surfaces. This 
effect blue-shifts quite slowly, as $a^{-2}$, so that it becomes rapidly irrelevant with respect to the unperturbed 
contraction: $k^2/(a^2 H^2) \sim t^{2-2p}$. Once the curvature has become negligible, a local observer will see a homogeneous, flat but 
anisotropic Universe: perturbations leaving the horizon have left an anisotropic initial condition. In synchronous gauge, the 
metric of an anisotropic homogeneous, flat space can be put in the form 
\be
ds^2 = -dt^2 + a^2(t) \sum_i e^{2 \beta_i (t)} dx^i dx^i \;, \qquad \sum_i \beta_i =0\;,
\ee
where $\beta_i(t)$ describe the anisotropic expansion. Their evolution is simply given by
\be
\dot\beta_i = c_i a^{-3} \; ,
\ee
and their effect in the Friedmann equation blueshifts as $a^{-6}$, 
\be\label{eq:aniF}
3 M_P^2 \, H^2 = 3 M_P^2 \,\left(\frac{\dot a}{a}\right)^2 = \rho + \frac1{2a^6} (c_1^2 + c_2^2 + c_3^2)  \; .
\ee
These equations together with the equation of motion for the scalar field, $\ddot\phi + 3H\dot\phi + V'(\phi)=0$, are
all we need to describe the anisotropic evolution of the Universe. It is straightforward to check that eqs.~(\ref{eq:syncg}) 
and (\ref{eq:syncphi}) are a linearized solution of the equations above. The given non-linear description
of an anisotropic Universe has been used in refs.~\cite{Heard:2002dr,Erickson:2003zm} to show that a contracting phase 
with $w > 1$ has an
attractor towards which nearby solutions flow. All curvatures and anisotropies get diluted away, the
fluctuations in the scalar drop to zero and we locally get back to the unperturbed solution up to exponentially small 
terms. In fact, even though it is not easy to get explicit solutions because the perturbation induces also fluctuations in the 
equation of state $w$, we see that the qualitative features we got in linear analysis remain valid. 
For instance in eq.~(\ref{eq:aniF}) the unperturbed energy density goes as $\rho \sim a^{-2/p}$, so that the anisotropic terms
give a relative correction going like $a^{-6+2/p} \sim t^{2-6p}$, which becomes irrelevant for $p< 1/3$. We
therefore expect that the unperturbed solution is recovered up to corrections that drop off like powers of $t^{1-3p}$, in
close analogy with the linear case.

Following the same argument we used in the previous section we can neglect the exponentially small terms. Although distant
regions of the Universe experience the same unperturbed history, as in the linear case their local metric is characterized by 
different, locally unobservable, integration constants. In general these constants 
describe the rescalings
of the three spatial coordinates (along axes that can rotate from point to point)
and thus the metric takes the form
\be \label{eq:NL}
ds^2 = -dt^2 +  e^{2 \zeta(\vec x)} e^{2 \gamma_{ij} (\vec x)} \, a^2(t) dx^i dx^j \;, \qquad \gamma_{ii}=0 \;.
\ee
The anisotropic piece $\gamma_{ij}$ comes from the fact that we have not restricted the analysis to scalar perturbations, as we instead 
did in the previous sections. Given a generic scalar-tensor initial configuration, going towards the bounce the system is driven to 
the form eq.~(\ref{eq:NL}) up to exponentially small terms.
The prescription to follow the long wavelength perturbations to the expanding phase is now clear: 
as in the linear case, the metric will have the same form eq.~(\ref{eq:NL}) after the bounce with the same space-dependent constants.
They cannot change between the contracting and expanding phase because their variation would imply observable deviations 
from the unperturbed solution, while we know that every comoving observer is following the unperturbed history up to exponentially
small terms.

In the long-wavelength approximation, {\em i.e.} for modes well outside the horizon, the variables $\zeta$ and $\gamma$ are 
{\em non-linearly} conserved independently of the phases the Universe goes through. This property is crucial to follow
the perturbations created during inflation and to make the non-linear observables, for instance the 3-point function of density
perturbations, independent of the unknown physics of reheating \cite{Salopek:1990jq,Maldacena:2002vr}. The same property 
holds also in our case, always assuming that the unknown bouncing phase is not sensitive to the exponentially small deviations 
from the unperturbed solution. Notice that the same assumption is implicitly made in the case of reheating.

\section{Conclusions}

The standard cosmological model has been extremely successful at explaining the evolution of the structures we observe 
in the Universe, from the early epochs we measure using the anisotropies in the cosmic microwave background to the large 
scale structures in the distribution of galaxies we see locally. However  this model requires a mechanism to create initial 
perturbations that correlate points with separations larger than the horizon. Such initial seeds will then grow under the 
action of gravity to form the structures we observe.

The natural assumption is to postulate that the initial seeds were created during an early phase in the history of the Universe 
and somehow ``stretched" outside the horizon. The ``stretching" of perturbations  can be accomplished in two different ways. 
One can have a period of accelerated expansion in which the length scale of perturbations grows rapidly, becoming larger than 
the horizon which evolves only slowly. This is done in inflationary models. The other possibility is to have the Universe 
contract slowly so that the horizon shrinks faster than does the scale of  perturbations. 
 
One fact is certain,  whatever the scenario, accelerated expansion or slow contraction, this period has to end and give rise to 
the standard hot Big Bang phase.   In inflationary scenarios the transition epoch is called reheating and could involve rather 
complicated physics. The beauty of inflation lays on the fact that the predictions of the model are independent of the details 
of reheating. The insensitivity to reheating stems from the fact that the wavelengths of interest to astronomy were extremely 
large compared to the horizon at the time and to the presence of a dynamical attractor. If the details of reheating were crucial 
to determining the predictions of inflation the scenario would not be that appealing. 

If perturbations were created during a contracting phase, with the Universe evolving towards a big crunch, the connection between 
the contracting phase and subsequent hot Big Bang phase may seem more problematic. The Universe has to go through a bounce with 
curvatures diverging as it approaches. This has been considered problematic by many detractors of such scenarios; in our mind 
however such criticisms are a bit unfair. Just as for inflationary models the important point is that their predictions are 
insensitive to the details of reheating, the real question for bouncing models is not how the bounce happened but whether or not 
one can show that their predictions are independent of the yet unknown UV physics responsible for the bounce. If one could argue 
that predictions are independent of the details of the bounce itself  then  bouncing models would be in almost as good a shape 
as inflation.

In this paper we have studied whether the predictions for density perturbations in the ekpyrotic/cyclic scenario are robust, 
{\em i.e.}~independent of the details of the unknown bouncing phase. The presence of a dynamical attractor in the contracting 
phase leads to a simple and natural assumption, namely that the physics of the bounce is independent of the exponentially small 
deviations from the unperturbed solution present as the bounce approaches. Even though seldom stated, this is the exact same 
assumption one makes in inflationary models about reheating. This assumption allowed us to calculate all the statistical 
properties of the fluctuations in the expanding phase. Unfortunately the spectrum is not scale
invariant and thus incompatible with the data. 

If this assumption is evaded, all predictions (scale invariance of the 
spectrum, level of non-gaussianities, etc.) strongly depend on the physics in the high curvature regime of the bounce. In this case, 
predictions cannot be made without a full resolution of the singularity. In a sense our conclusions are more pessimistic than just 
the statement that in order to make bouncing models viable one needs to understand the details of the bounce.  We have shown that 
from the perspective of a local observer the bounce has to be exponentially sensitive to the initial state.  

In the literature many prescriptions have been proposed to match the two independent modes in the contracting phase to those 
in the expanding one \cite{Lyth:2001pf,Brandenberger:2001bs,Hwang:2001zt,Martin:2001ue, Durrer:2002jn,Martin:2004pm}. These 
prescriptions usually involve variables (like the Newtonian potential $\Phi$) which diverge going towards the bounce and thus 
hide both the 
perturbativity of fluctuations and the existence of a dynamical
attractor, features which are instead manifest in the synchronous gauge. Moreover, as different 
variables like $\Phi$ and $\zeta$ behave very differently in the contracting phase, one is lead to different ``natural'' prescriptions 
depending on which variable and gauge is used. We think that these mathematical prescriptions are not based on well motivated 
physical assumptions, and we consider the attractor as the only physical guide across the bounce. In fact it should be now clear 
that prescriptions giving a scale invariant spectrum  describe a bouncing phase which is sensitive to exponentially small 
departures from the unperturbed solution, or in which all relevant modes come back into the horizon. In both cases results 
are anyway completely dependent on the unknown UV physics. 

For example some of the proponents of the ekpyrotic/cyclic scenario have developed a matching
procedure based on analytic continuation in the full 5-d setup \cite{Tolley:2003nx} (see also ref.~\cite{Battefeld:2004mn}). 
The result is that $\zeta$ has a jump at the bounce proportional to the comoving energy density perturbation $\epsilon_m$, 
$\Delta\zeta \sim \epsilon_m/(k_{\rm ph} L)^2$, where $L$ is a UV length scale characterizing the physics of the bounce. 
The variable $\epsilon_m$ goes to zero during the contracting
phase as $t^{1-3p}$ and it is therefore exponentially small at the bounce. Therefore from the 4d perspective this prescription describes
a bounce which is exponentially sensitive to the incoming state; $\epsilon_m$ is in fact multiplied by a huge $k$-dependent
term. This extreme sensitivity to initial conditions is clearly problematic; for instance it amplifies also all preexisting
anisotropies which were diluted away during the contracting phase thanks to the presence of the attractor. 

We conclude stressing that the situation is slightly different for pre-Big Bang scenarios \cite{Gasperini:2002bn}. In this case there
is no potential for the scalar (dilaton) and only its kinetic energy is driving the (Einstein frame)
contracting phase; this corresponds to the limit $p \to 1/3$ in our solutions of section \ref{intuitive}. The kinetic energy 
blue-shifts as $a^{-6}$, exactly like the contribution of anisotropies to the Friedmann equation (\ref{eq:aniF}). This implies
that we are somewhat borderline with respect to the existence of the attractor. In fact the anisotropic
perturbation in synchronous gauge and the fluctuation in the scalar field, instead of going to zero, diverge logarithmically. 
The situation now resembles an inflationary cosmology in the presence of isocurvature 
perturbations; separated regions of the Universe are characterized by a different mixture of two independent modes and
therefore go through the bounce in a different way depending on the relative size of the two solutions. Our conclusions
cannot be applied and it is necessary to specify the details of the high curvature regime to evolve the fluctuations across
the bounce.

\section*{Acknowledgments}
We thank Nima Arkani-Hamed, Massimo Porrati, Leonardo Senatore, Toby Wiseman and especially Justin Khoury and Paul Steinhardt for useful discussions. 
M.~Z.~ is supported by NSF grants AST 0098606 and by the David and Lucille Packard Foundation Fellowship for Science and Engineering
and by the Sloan Foundation.

\appendix

\section*{Appendix}

\section{The quadratic action for scalar fluctuations}\label{app_action}
The complete action for gravity and a minimally coupled scalar field $\phi$ is
\be \label{complete_action}
S= \half \int \! d^4 x \sqrt{-g} \, \left[ R - (\partial_\mu \phi)^2 - 2V(\phi) \right] \; .
\ee
(For notational simplicity we are setting $M_P =1$; at the end of the computation $M_P$ can be restored by dimensional 
analysis).
In order to find an action for the scalar fluctuations of this coupled system around a given background
it is convenient to work in the ADM formalism. This has been done in ref.~\cite{Maldacena:2002vr}, and in this 
section we briefly review the results derived there.
In terms of the ADM variables the metric is
\be
ds^2 = -N^2 dt^2 + h_{ij}(dx^i + N^i dt)(dx^j + N^j dt) \; ,
\ee
and the action eq.~(\ref{complete_action}) reads
\be
S= \half \int \! d^4 x \sqrt{h} \, \left[ 
N R^{(3)} - 2NV(\phi) + N^{-1}\big(E_{ij} E^{ij} - E^2\big) 
+ N^{-1}\big(\dot \phi - N^i \partial_i \phi\big)^2
- N h^{ij} \partial_i \phi \partial_j \phi  \right] \; .
\ee
Here $R^{(3)}$ is the intrinsic curvature of spatial slices, while $E_{ij}$ is related to the extrinsic curvature,
\be
E_{ij} = \half\big( \dot h_{ij} - \nabla_i N_j - \nabla_j N_i \big) \; , \qquad K_{ij} =  N^{-1} E_{ij} \; ,
\ee
and $E = E^i{}_i$.
From the action above it is clear that $N$ and $N^i$ are just Lagrange multipliers. One can then solve for them 
the constraint equations, express them in terms of the other degrees of freedom,
and substitute their value back in the action.
At this stage it is useful to pick a gauge. We choose the $\zeta$-gauge, defined at linear order in the fluctuations by
\be
\phi(\vec x, t) = \phi_0 (t) \; , \qquad h_{ij} = a^2(t)\big(1+2\zeta \big) \delta_{ij} \; ,
\ee
where $\phi_0 (t)$ and $a(t)$ are the background solutions for the scalar and the scale factor, and we have neglected 
tensor modes.  
In this gauge the constraint equations read
\begin{eqnarray}
\nabla_i \left[ N^{-1} (E^i {}_j - \delta^i_j E)\right] & = & 0  \; , \\
R^{(3)} - 2 V(\phi_0) - N^{-2}\big(E_{ij} E^{ij} - E^2\big) - \dot \phi_0^2 & = & 0  \; ,
\end{eqnarray}
whose solution at linear order is
\be
N = 1 + \frac{\dot \zeta}{H}\; , \qquad \qquad N^i = \partial_i \Big( - \frac{\zeta}{a^2 H} + \chi \Big)\;, 
\quad \partial_i \partial_i \chi = \frac{\dot \phi_0^2}{2 H^2} \dot \zeta \; .
\ee
Plugging this into the action above and expanding up to second order one gets a quadratic action for the 
only scalar degree of freedom $\zeta$.
After integrating by parts and using the background equations of motion ({\em i.e.}~the Friedmann equation and the equation 
for the scalar) one finally gets
\be
S=\half \int \! dt d^3 x \, \frac{\dot \phi_0^2}{H^2}\left[ a^3 \dot \zeta^2 - a (\partial_i \zeta)^2 \right] =
- \half \int \! d^4 x \sqrt{g} \, \frac{\dot \phi_0^2}{H^2} \, ( g^{\mu\nu}\, \partial_\mu \zeta \partial_\nu \zeta)  \; ,
\ee
where $g_{\mu\nu}$ is the background FRW metric.
For the scaling solution discussed in the text the quantity $\dot \phi_0^2/H^2$ is a constant, and the action above
reduces simply to eq.~(\ref{action_zeta}). 

In the alternative gauge eq.~(\ref{gauge_deltaphi}), which keeps $\delta\phi$ as the scalar dynamical variable, 
the result for the action is not as simple as in the $\zeta$-gauge. 
This is because the relation between $\zeta$ and $\delta\phi$ is in general time-dependent, 
$\zeta = -(H/\dot \phi_0) \, \delta\phi$, so that time derivatives acting on $\zeta$ in the action above also produce
time derivatives of the factor $H/\dot \phi_0$. 
However, for the scaling solution we are interested in such a factor is time-independent, so that again the quadratic
action takes the very simple form eq.~(\ref{action_deltaphi}).

\section{From $\zeta$-gauge to synchronous gauge}\label{app_gauge}
In this section we explicitly derive the gauge transformation that goes from $\zeta$-gauge to synchronous gauge.
In $\zeta$-gauge the metric for a general scalar fluctuation is at linear order (see the previous section)
\begin{eqnarray}
g_{00} & = & -1-2\frac{\dot \zeta}{H} \; ,\\
g_{0i} & = & - \partial_i \bigg[\frac{\zeta}{H} -  \frac{\dot\phi_0^2}{2H^2}\, \frac{a^2}{\partial^2}\dot\zeta \bigg] 
\equiv  - \partial_i \Xi \; , \label{Xi}\\
g_{ij} & = & a^2 \big(1+2\zeta\big) \delta_{ij} \label{gij}\; .
\end{eqnarray}
Under a gauge transformation of parameter $\xi_\mu$ the metric transforms as $g_{\mu\nu} \to g_{\mu\nu} 
+ \nabla_\mu \xi_\nu + \nabla_\nu \xi_\mu$, where the $\nabla$'s are the covariant derivatives associated to the
unperturbed FRW metric. The non-zero Christoffel symbols in a FRW geometry are
\be
\Gamma^0_{ij} = a^2 H \, \delta_{ij} \; , \qquad \Gamma^i_{0j} = H \delta^i_j \; ,
\ee
so that the explicit transformation of the metric components is
\begin{eqnarray}
g_{00} & \to & g_{00} + 2 \dot \xi_0 \; , \\
g_{0i} & \to & g_{0i} + \dot \xi_i + \partial_i \xi_0 - 2H \xi_i = 
g_{0i} + a^2 \frac{d}{dt} \frac{\xi_i}{a^2} + \partial_i \xi_0 \; , \\
g_{ij} & \to & g_{ij} + \partial_i \xi_j + \partial_j \xi_i - 2 a^2 H \delta_{ij} \xi_0  \label{trans_gij}\; . 
\end{eqnarray}
We want to end up in synchronous gauge, which is defined by $g_{00} = -1$ and $g_{0i} = 0$. The explicit form of $g_{00}$ 
combined with its transformation law tells us that we must choose
\be
\dot \xi_0 = \frac{\dot \zeta}{H} \quad \Rightarrow \quad \xi_0 = \int_0^t \! \frac{\dot \zeta}{H} dt' \; .
\ee
To be completely general we should also add to $\xi_0$ a generic function of $\vec x$, 
but it will turn out that it is not necessary, so we set it to zero.

Then we want to set to zero the space-time components $g_{0i}$. In order to do so $\xi_i$ must satisfy
\be
\frac{d}{dt} \frac{\xi_i}{a^2} = \frac{1}{a^2} \partial_i \bigg[ \Xi - \int_0^t \! \frac{\dot \zeta}{H} dt' \bigg] \; ,
\ee
so that
\be
\xi_i = a^2 \partial_i \int_0^t \!dt' \frac{1}{a^2}\bigg[ \Xi - \int_0^{t'} \! \frac{\dot \zeta}{H} dt'' \bigg] \; .
\ee
The last step is to combine eqs.~(\ref{gij}, \ref{trans_gij}) and read off the spatial metric in synchronous gauge,
\be
g_{ij} \to  a^2 \left\{ \Big( 1+2\zeta - 2H \int_0^{t} \! \frac{\dot \zeta}{H} dt' \Big) \delta_{ij}
+2 \partial_i \partial_j \int_0^t \!dt' \bigg[ -  \frac{\dot\phi_0^2}{2H^2}
\frac{1}{\partial^2} \dot\zeta
+\frac{1}{a^2}
\Big(\frac{\zeta}{H} - \int_0^{t'} \! \frac{\dot \zeta}{H} dt'' \Big)\bigg]
\right\} \; ,
\ee
where we used the definition of $\Xi$, eq.~(\ref{Xi}).
Specializing to our scaling background $a \propto t^p$, $H=p/t$, $\dot\phi_0 = \sqrt{2p}/t$ we get precisely eq.~(\ref{eq:syncg}).

Finally we want to express the scalar $\phi(\vec x, t)$ in synchronous gauge. Under a gauge transformation
\be
\phi(x) \to \phi(x) + \partial_\mu \phi(x) \cdot \xi^\mu \; .
\ee
Since in $\zeta$-gauge the scalar is unperturbed, $\phi(\vec x, t) = \phi_0(t)$, in synchronous gauge we simply have
\be
\delta\phi \equiv \phi(\vec x, t) - \phi_0(t) = \dot \phi_0  \, \xi^0 = - \dot \phi_0  \, \xi_0 
= - \dot \phi_0 \int_0^t \! \frac{\dot \zeta}{H} dt' \; .
\ee
For our scaling solution this reduces to eq.~(\ref{eq:syncphi}).

\end{document}